\documentstyle[11pt,newpasp,twoside,epsf]{article}
\def\edcomment#1{\iffalse\marginpar{\raggedright\sl#1\/}\else\relax\fi}
\reversemarginpar

\begin{document}
\title{High energy astrophysical neutrinos}
 \author{H. Athar}
\affil{Physics Division, National Center for Theoretical Sciences,
 101 Section 2, Kuang Fu Road, Hsinchu 300, Taiwan}

\begin{abstract}
 High energy neutrinos with energy typically greater than tens of
 thousands of GeV may originate from several astrophysical sources.  The sources may
 include, for instance, our galaxy, the active centers of nearby
 galaxies, as well as possibly the distant sites
 of gamma ray bursts. I briefly review some aspects of production and
 propagation as well as prospects for observations of these
 high energy astrophysical neutrinos.
\end{abstract}

\section{Introduction}

During\footnote{Talk given at IAU 8th Asian Pacific Regional
Meeting, 2$-$5 July, 2002, Tokyo, Japan.} nearly past 50 years,
the empirical search for neutrinos has spanned roughly six orders
of magnitude in energy, from approximately $10^{-3}$ GeV up to
approximately $10^{3}$ GeV. The lower energy edge corresponds to
solar neutrinos, whereas the upper energy edge corresponds to
atmospheric neutrinos. The intermediate energy range include the
terrestrial and supernova neutrinos. This search has already given
us remarkable insight into neutrino interaction properties as well
as its intrinsic properties such as mixing and mass. Here, I
briefly review the possibility of having neutrinos with energy
greater than $10^{3}$ GeV. The upper energy edge for high energy
astrophysical neutrinos is limited only by the concerned
experiments. A main motivation to search for such high energy
astrophysical neutrinos is to get more accurate information about
the origin of observed high energy photons (and ultra high energy
cosmic rays) that is presently not possible through conventional
gamma ray astronomy. For instance, the observation of high energy
gamma ray flux alone from active centers of nearby galaxies (AGNs)
such as M 87 and distant sites of gamma ray bursts (GRBs) does not
allow us to identify its origin in purely electromagnetic or
purely hadronic interactions unambiguously. Sizable high energy
astrophysical neutrino flux is expected if latter interactions are
to play a dominant role. Search for high energy astrophysical
neutrinos will thus provide us a complementary and yet unexplored
view about some of the highest energy phenomenons occurring in the
known universe. For a general introduction of the subject of high
energy astrophysical neutrinos, see Bahcall \& Halzen (1996),
Protheroe (1999), Bahcall (2001). See, also Battiston (2002).

\section{High energy astrophysical neutrinos}
\subsection{Production}

The purely hadronic (such as $pp$ or $pn$) and photo hadronic
(such as $\gamma p$ or $\gamma n$) interactions taking place in
cosmos currently represent the main source interactions for the
production of high energy astrophysical neutrinos. Examples of the
astrophysical sites where these interactions (may) take place
include our galaxy, the AGNs and the GRBs. In some model
calculations for high energy astrophysical neutrino flux, the
proton acceleration mechanism is considered to be the same as for
electron acceleration at the astrophysical sites.

The accelerated protons in the above interactions in these sites
produce unstable hadrons such as $\pi^{\pm}$ and  $D_{s}^{\pm}$
that decay into neutrinos of all three flavors. The same
interactions also produce $\pi^{0}$ that can contribute dominantly
towards the observed high energy photons, whereas the escaping
accelerated protons may (or may not) dominantly constitute the
observed ultra high energy cosmic rays depending upon the finer
details of the relevant astrophysical site. The absolute
normalization of the high energy astrophysical neutrino flux is
obtained by assuming that a certain fraction of the observed high
energy photon flux has (purely) hadronic origin and (or) that the
observed  ultra high energy cosmic ray flux can dominantly
originate from that class of astrophysical sites. Typically, the
muon neutrino flux is twice the electron neutrino flux with
essentially negligible tau neutrino flux at the production site.
For a recent review article, see Halzen \& Hooper (2002). High
energy astrophysical neutrino production is in principle also
conceivable in purely electromagnetic (such as $\gamma \gamma $)
interactions taking place in cosmos, see Athar \& Lin (2002).

The dedicated high energy neutrino detectors provide us a clue as
well as a check for the absolute normalization of the high energy
neutrino flux. For instance,  the present upper bound from
Antarctic Muon and Neutrino Detector Array (AMANDA)  give value of
$9.8\cdot 10^{-6}$ cm$^{-2}$ s$^{-1}$ sr$^{-1}$ GeV  for absolute
flux of diffuse high energy neutrinos for the energy range between
$5\cdot 10^{3}$ GeV to $3\cdot 10^{5}$ GeV, see
 Ahrens et al. (2002). The AMANDA (B10) is at the south
 pole and its current upper bound is based on non observation.
 Its present cylindrical configuration searches for upward
 going high energy (muon) neutrinos covering the northern
 hemisphere with an effective area of $\sim 0.01 $ km$^{2}$ for
 (muon) neutrino energy $\sim 10^{4}$ GeV.

\subsection{Propagation}

    With three light stable neutrinos, as suggested by
    standard model of particle physics, neutrino flavor mixing is a
    dominant effect during high energy astrophysical neutrino propagation, once
    they are produced, see Athar, Je\.{z}abek, \& Yasuda (2000).
    Since the average interstellar matter density
    is rather low, therefore the neutrino nucleon deep inelastic scattering (DIS)
    effects are usually negligible. These neutrinos thus restore the arrival direction and
    energy information starting from the production site. On the other hand, because of
    rather large unobstructed distances traversed by these neutrinos, typically
    greater than 1 pc, where 1 pc $\simeq 3\cdot 10^{18}$ cm, the neutrino flavor
    mixing equally distribute the three neutrino flavors in the
    mixed
    high energy astrophysical neutrino flux.
    The present neutrino oscillation data implies that the deviations from
    these symmetric distributions are not more than a few percent. The
    absolute level of (downward going) high energy astrophysical
electron
    neutrino flux, arriving at the detector,  is essentially
    independent of the relative ratio at the
    production, as it is least effected by neutrino oscillations,
    see Athar \& Lin (2001).

\subsection{Prospects for observations}

    Typical high energy astrophysical neutrino observation can be achieved by attempting
    to observe the Cherenkov radiation from the associated charged
    leptons and/or showers produced in DIS occurring near or inside
    the detector. For simplicity, here I ignore a possible observational
    difference between neutrinos and anti neutrinos.

    The mixed high energy astrophysical neutrino flux arrives at
an earth based detector in three general directions. The downward
going
    neutrinos do not cross any significant earth cord while reaching the detector.
    In Fig. 1, the downward going
    event
    rate for the three neutrino flavors is displayed under the
\begin{figure}
\plotone{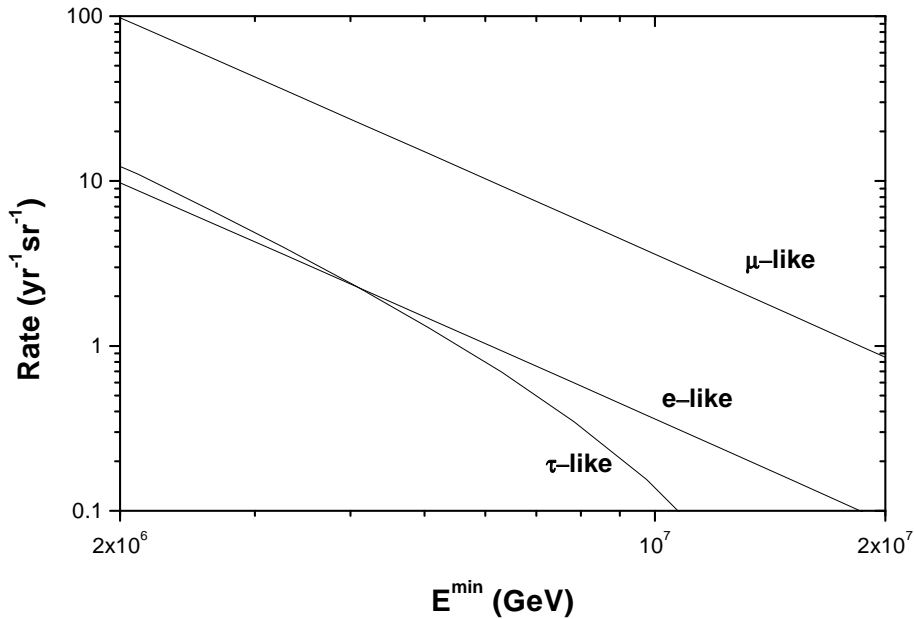} \caption{Expected downward going $e-$like,
$\mu-$like and $\tau-$like event rate produced by AGN neutrinos as
a function of minimum energy of the corresponding charged lepton
in a large km$^{3}$ volume ice or water neutrino detector. Three
flavor neutrino mixing is assumed.}
\end{figure}
  assumption of neutrino flavor mixing in units of yr$^{-1}$ sr$^{-1}$
    as a function of minimum energy of the corresponding charged lepton
    produced in charged current DIS occurring near or inside the detector.
    The different levels and energy
    dependencies of the rates reflect the differences in the
    associated charged lepton ranges,
    in a large (under construction)  km$^{3}$ volume ice or water
    detector such as the proposed extension of AMANDA (B10), although
    as mentioned in the last subsection, the three
    neutrino flavors arrive at the detector in almost equal proportion.
      The contained $e-$like event rate is obtained by
    rescaling the $\mu-$like event rate for illustration. A diffuse AGN neutrino flux model
    is used here as an example where $pp$ interactions
    are considered to play an important role, see Szabo \& Protheroe (1994).
    The event topology for each flavor can
    possibly be identified separately in  km$^{3}$ volume water or ice detector
    within the energy range shown in Fig. 1. For details, see Athar, Parente, \& Zas (2000).

    The near horizontal neutrino
    flux  crosses a small earth cord before reaching the
    detector. Several proposals are under study to construct a
    specific detector for such type of neutrinos, see Hou \& Huang (2002).
    The upward going neutrino flux
    crosses a significant earth cord before reaching the
    detector, and is therefore absorbed by the earth to a large extent
    for energy typically greater than 5$\cdot 10^{4}$ GeV, see, for instance,
    Hettlage \& Mannheim (2001).
    Above this energy,
    the earth diameter exceeds the charged current DIS length.
    In addition to attempting to measure  the Cherenkov radiation, several other
    alternatives are also presently being explored. See, for instance,
    Chiba et al. (2001).

\section{Conclusion}

    Several astrophysical sources such  as center(s) of our
    as well as other nearby galaxies may produce high
    energy astrophysical neutrino flux with energy around or above
     $10^{4}$ GeV, compatible with the observed high energy
     gamma ray and ultra high energy cosmic ray flux
     above the atmospheric neutrino flux. This
     implies need for a more meaningful search.

        The author thanks Physics Division of National Center for
        Theoretical Sciences for financial support.

\end{document}